\documentclass[12pt, preprint]{aastex}
\usepackage[]{natbib}
\usepackage{graphics}
\usepackage{amssymb}
\usepackage{amsmath}

\slugcomment{} \shorttitle{CaII absorption in SDSS J2339$-$0912}
\shortauthors{Wang et al.}
\slugcomment{ApJL accepted}

\begin{document}

\title{ Strong CaII absorption lines in the reddened quasar SDSS J2339$-$0912:
    Evidence of the collision/merger in the host galaxy? }

\author{T.G. Wang, X.B. Dong, H.Y. Zhou and J. X. Wang}
\affil{Center for Astrophysics, University of Science
and Technology of China, Hefei, Anhui, P.R.China}
\email{twang@ustc.edu.cn}

\begin{abstract}
We report the detection of strong CaII/MgII absorption lines at
the quasar redshift in the narrow line quasar SDSS J2339$-$0912 (
$z=0.6604$). The quasar exhibits strong FeII, small Balmer
emission line width and a very red $B-K_{s}$ color. Both the
optical continuum and broad emission lines are reddened by
SMC-like dust of E(B$-$V)$\simeq$1.0 mag, while its near-infrared
color ($J-K_{s}=1.60$) shows little reddening. The CaII absorption
lines are saturated and resolved with an FWHM of 362 km~s$^{-1}$
and an equivalent width of $W_{\mathrm CaII\; K}=4.2$\AA~ (in the
source rest frame). MgII absorption lines are also saturated and
have a similar line width. The line profile and the fact that
there is no evidence for starlight from the host galaxy suggest
that these absorption lines are not of a stellar origin. The ratio
of column density of CaII to that of dust is consistent with that
of the ISM in our Galaxy. We suggest that both the heavy reddening
and the large absorption line width are due to the highly
disturbed ISM on the line of sight toward the quasar, and that the
disturbance is caused by a galaxy collision or even merger in the
quasar host galaxy.

\end{abstract}


\section{Introduction}

Quasars are characterized by blue continuum and strong broad
emission lines. However, a fraction of them show red colors due to
wavelength-dependent extinction by dust on the line of sight. In
the past years, we have witnessed controversial evidence on
whether a large population of such objets exist or not (Cutri et
al. 2000; Webster et al. 1995; Richards et al. 2003). The
absorption can be caused by the dusty material either in the
nucleus (i.e., inside the Narrow Line Region [NLR]), in the host
galaxy or in the intervening galaxies (Pei, Fall \& Bechtold
1991). Recent studies of red quasars from the SDSS (the Sloan
Digital Sky Survey; York et al. 2000) suggested that most quasars
are generally reddened by dust at the quasar redshifts (Richards
et al. 2003; Hopkins et al. 2004), whether it is in the host
galaxies or near the nucleus. Evidence for substantial absorbing
material within the NLR is obvious in type 1.8/1.9 quasars
selected from the SDSS, since their broad lines are heavily
reddened while narrow lines not (Dong et al. 2005). Potential use
of these reddened quasars in the study of the host galaxy
properties has been discussed by these authors.

Reddened quasars are also important in studying the gas/dust
environment of quasars and AGNs. Extinction curve is an important
element to characterize the reddening in individual quasars. In
the past years, there is still a controversy regarding what form
of extinction curve should be applied to the quasar's intrinsic
reddening (e.g., Czerny et al. 2004; Hopkins et al. 2004).

In this paper, we report the detection of strong CaII and MgII
absorption lines at the quasar redshift in the heavily reddened
narrow line quasar SDSS J233903.82$-$091221.2 ($z$=0.6604; hereafter
SDSS J2339$-$0912), which was initially discovered by the FIRST
bright quasar survey (FBQS J2339$-$0912; Becker et al. 2001) with a
moderate radio flux 4.3 mJy at 21~cm. We picked it up as an
unusual Narrow Line Seyfert 1 galaxies (NLS1s) during the
statistical study of a sample of NLS1s selected from the SDSS Data
Release 2 and 3 (DR2 and DR3; Abazajian et al. 2004 \& 2005). We study the
absorption lines as well as the continuum and emission lines of
this object, and suggest that the heavy reddening and the large
absorption line width are due to the highly disturbed ISM of the
host galaxy on the line of sight toward the quasar.

\section{Modelling the continuum, absorption lines and emission lines}

SDSS J2339$-$0912 is unusual for its very red spectrum in the sample
of $\sim$ 1500 NLS1s selected from the SDSS DR2+DR3 spectroscopic
catalogs. The SDSS optical spectrum, with a total exposure time of
3788 sec, is shown in Fig 1. CaII K, H absorption lines are
evident. Due to its red color, S/N ratio drops rapidly towards
short wavelength, and reaches as low as unity below 4100 \AA~
(2500 \AA~ in the source rest frame). Nonetheless, MgII doublet
absorption lines are visible, while the corresponding emission
lines are hardly detected. At long wavelengths, the spectrum shows
prominent FeII and H$\beta$ lines. [OIII] $\lambda$5007 emission
line, blended with Fe II multiple 42, is visible and weaker than
[OII]$\lambda$3727.

The spectrum was brought to the rest frame using the redshift
determined by [OII]$\lambda$3727, and corrected for Galactic
reddening of E(B$-$V)=0.027 mag. The continuum and FeII emission
lines are modelled 
as follows,
\begin{equation}
f(\lambda)=A(E_{B-V},\lambda)[bB(\lambda)+cC(\lambda)]
\end{equation}
where $B(\lambda)$ is the FeII templates obtained by Veron-Cetty
\& Veron (2003) covering the wavelengths between 3535$-$7534 \AA,
and $C(\lambda)=(\lambda/4000$\AA$)^{-1.7}$ the power-law
continuum (Francais 1996). We assume that FeII has a profile of
Lorentzian, as Balmer lines (see below). We do not include UV FeII
multiplets in the fit because the S/N ratio at the blue part of
the spectrum is very low. For the same reason, we do not include
the model of Balmer continuum. Adopting a single power-law
prescription for the quasar continuum should be reasonable
considering the low S/N ratio of the spectrum at the blue end. The
best fitted model is obtained by minimizing $\chi^2$. Emission
lines apart from the FeII and absorption lines were masked out
during the fitting. We have tried different extinction curves, but
only SMC-like curve can give a good fit while others lead to much
worse result (see next section). The final fit is also shown in
Fig 1. This simple model reproduces quite well the overall
observed continuum and FeII spectrum. The best fitted E(B$-$V) is
0.97 mag for SMC-like dust.

Emission and absorption line parameters are measured after the
continuum and FeII emission being subtracted. H$\beta$ is fitted
with a Lorentzian; [OII] and [OIII] lines Gaussians. These models
can fit the observed profile quite well. The width of H$\beta$
(2194$\pm$108 km~s$^{-1}$) and the reddening corrected
FeII/H$\beta$ ratio ($R_{4570}\equiv
FeII\lambda\lambda$4434-4684$/H\beta=1.20$) are consistent with
the definition of NLS1s (Goodrich 1989). With a
reddening-corrected luminosity $\lambda L_\lambda
(5100)=2\times10^{46}$ erg s$^{-1}$, it undoubtedly is a reddened
narrow line quasar. The [OII]/[OIII] ratio is 1.65$\pm$0.32. But
[OIII] line profile is affected by red noise in the spectrum due
to the sky lines, and thus its profile is determined less
reliably.

Gaussian absorption line profile is used to fit the CaII K, H
doublet. The widths of the two lines are set to be identical
during the fit. Given the spectral resolution, it is not
surprising that this model gives an acceptable fit. The equivalent
width (EW) of CaII K line is 4.20$\pm$0.24 \AA~in the source rest
frame; its width is 362$\pm$34 km~s$^{-1}$ FWHM, corrected for the
instrumental broadening of 170 km~s$^{-1}$. The observed ratio of
the EWs of K to H lines, $W_{ {\mathrm CaII\; K}}/W_ {{\mathrm
CaII\; H}}$, is $1.35\pm 0.13$, indicating the saturation. MgII
doublet absorption lines are saturated also and have widths
similar to CaII doublet. But they cannot be measured exactly due
to the low S/N ratio of the data. The derived emission and
absorption line parameters are listed in Table 1.

\begin{table*}
\begin{center}
{\bf Table 1.} Emission line and absorption line parameters of SDSS J2339$-$0912. \\[0.1cm]
\nobreak
\begin{tabular}{lrrrr}
\hline \hline
Line & Centroid\tablenotemark{a} & W\tablenotemark{a} & FWHM\tablenotemark{a} \\
 & \AA & \AA & km~s$^{-1}$ \\
\hline
$[OIII]\lambda 5007$ & 5005.9$\pm $0.4 & 2.5$\pm 0.1 $ & 372$\pm $31 \\
$H\beta $ & 4858.9$\pm $0.5 &  39.5$\pm $1.5  & 2194$\pm $108 \\
FeII$\lambda \lambda $4570 & & 44.6 &  \\
CaII H & $3966.7\pm 0.2$ & $3.1\pm 0.2$ & $400\pm34$ \\
CaII K & $3932.1\pm 0.2$ & $4.2\pm 0.2$ & $400\pm34$ \\
$[OII]\lambda 3727$ & 3728.5$\pm $0.3 & 8.8$\pm 0.1 $ & 310$\pm$24\\

\hline
\end{tabular}
\tablenotetext{a}{~Line centroid and equivalent width are given in
the source rest frame. FWHM has not been corrected for the
instrumental broadening.}
\end{center}
\end{table*}

\section{Discussion}

\subsection{The Extinction Curve of Dust Reddening}

The extinction laws in the quasar and Seyfert galaxies have become
an important topic in the past years. By combining the column
density derived from the X-ray absorption and the UV/optical
reddening, Maiolino et al. (2001) found that the $A_V/N_H$ for
Seyfert galaxies is significantly lower than the Galactic one by a
factor ranging from 3 to 100 (also Loaring, Page \& Ramsay 2003;
but Weingartner \& Murray 2002 and Carrera, Page \& Mittaz 2004
for controversy). The lack of prominent absorption feature at 9.7
$\mu$m and 2175 \AA~ leads the former authors to propose that
large grains are responsible for the reddening, with a flat
extinction curve in the UV. By comparing the composite spectra
derived from samples of radio selected quasars with different
orientations, Gaskell et al (2004) derived an extinction curve
much flatter than the Galactic one in the UV. However,  it was
found that SDSS red quasars have colors consistent with
significant reddening by an extinction curve similar to that of
SMC (Richards et al. 2003; Hopkins et al. 2004). Strong evidence
shows that the extinction curve is even steeper than that of SMC
in two unusual BAL QSOs with continuum reddening $E(B-V)\simeq$
0.5 (Hall et al. 2002a). Consistent with the latter, Zuo et al.
(1997) determined the reddening by comparing the spectra of two
images of gravitational lensed QSO 0957+561 with different
reddening. They derived a high dust-to-gas ratio and an extinction
curve steeper than the Galactic one in the UV. Based on the SDSS
composite reddened quasars, Czerny et al. (2004) proposed the
extinction can be modelled by an AC sample of amorphous carbon
grains.

Both the optical continuum and the broad emission lines in SDSS
J2339$-$0912 are heavily reddened by SMC-like dust of E(B$-$V)$\simeq$
1.0 mag at the redshift of the quasar. We have applied the
extinction curves of SMC, LMC and our Galaxy as well as those
derived by Gaskell et al (2004) and Czerny et al (2004) to model
the continuum and FeII spectrum of the quasar. We found that only
the SMC one can reproduce the observed spectrum well while others
cannot due to their flatness in the UV. This is shown in Fig 1. In
fact, at rest wavelengths below 8000\AA~, this object is the
reddened version of a normal narrow line quasar SDSS
J080908.14+461925.6 ($z$=0.6563; hereafter SDSS J0809+4619) by
SMC-like dust with E(B$-$V)=1.0 (see Fig 2). SDSS J0809+4619 is
selected by its blue colors, strong FeII emission and the
closeness to SDSS J2339$-$0912 in redshift. However, the observed
near-infrared $J-K_{s}$ colors (1.63 for SDSS J2339$-$0912 and 1.66
for SDSS J0809+4619 from the 2MASS) of the two objects are almost
the same, indicating no reddening of SDSS J2339$-$0912. The
inconsistent results drawn from the optical to near-UV spectrum
and from the near-infrared photometry are likely due to the
intrinsic difference in the near-infrared emission of the two
objects. First, the NIR SED of PG quasars shows large scatter, and
seems not to be correlated with the SED in the optical and UV
(Neugebauer et al. 1987). Especially, the reddened SED of PG
2344+092 ($z=0.677$) approximately matches the SED of SDSS
J2339$-$0912 (Fig 2). Second, if this inconsistence is interpreted as
different extinction laws in the near-infrared, a
wavelength-independent extinction curve over a wavelength range
from $\sim$ 8000\AA~ to 1.3$\mu$m is required, which is not
consistent with any known extinction curve. Finally, variability
is not likely responsible for the discrepancy because the J-band
flux of SDSS J2339$-$0912 is fairly consistent with the reddened one
of SDSS J0809+4619.

Although the reddening of the BLR is evident, we cannot infer the
reddening of the NLR. The EW of [OIII]$\lambda$5007\AA~ is already
in the lower end among PG quasars. If [OIII] line is not reddened
while the continuum is with E(B$-$V) $\approx$ 1.0 mag, the
intrinsic EW of [OIII]$\lambda$5007\AA~ would be 0.2\AA, at the
extreme.  On the other hand, the large [OII]$\lambda$3727\AA~ to
[OIII]$\lambda$5007\AA~ ratio suggests reddening is not important
for the NLR, if [OII] and [OIII] originate from the same region.

\subsection{The Origin of the Absorbing Material}

CaII absorption lines in this quasar is not of a stellar origin,
because there is no other feature indicative of old stellar
populations in the spectrum such as 4000\AA~ break. The presence
of saturated CaII and MgII absorption lines and the well-fitted
Gaussian profile of CaII lines suggest that the absorption is
caused by warm diffuse gas on the line of sight toward the quasar.
Thick cold dusty material is also present on the line of sight
according to the heavy reddening of both the continuum and the
broad emission lines. It should to be noted that there are large
residuals under the CaII absorption lines, suggesting that either
(1) the absorber partially covers the source, or (2) it is
constituted of several optically thick clouds. Partial covering is
not important if the warm gas co-exists with the dusty material as
we argue below, because the attenuation of the UV flux by
reddening is extremely large. In the case (2), a lower limit on
the column density can be imposed if we assume that the absorption
line is constituted of several equally saturated, unresolved
components. Using the growth curve of the absorption lines, we
obtain a lower limit on the column density
$\sim$5$\times$10$^{13}$~ cm$^{-2}$ from the ratio of the CaII
doublet and the EWs. The exact value can only be obtained through
high resolution spectroscopic observation.

With our estimated (lower limit on the) CaII column density, the
ratio of N(CaII)/E(B$-$V) is very close to those observed in the
Galactic or toward Magllanic clouds. Wakker \& Mathis (2000) found
that HI column density is correlated with the column density of
CaII with standard deviation of 0.40 dex for high- and
intermediate velocity clouds in the Galaxy. More recently, Smoker
et al. (2003) studied CaII absorption on the lines of sight to 88
mainly B-type stars, and found that log[N(CaIIK)/N(HI)] ranges
from $-$7.4 to $-$8.4, 0.5 dex higher than that of Wakker \&
Mathis (2000). For the Galactic clouds, Bohlin, Savage, \& Drake
(1978) found N(H I)/E(B$-$V)$= 4.8\times 10^{21}$~ cm$^{-2}$,
which gives $13.3< \log N(CaII~K)/E(B-V) < 14.3$, consistent with
the value obtained in SDSS J2339$-$0912. Thus, we propose that the
warm gas responsible for the CaII/MgII absorption in SDSS
J2339$-$0912 co-exists with the dusty material responsible for the
reddening, with a gas-to-dust ratio similar to the ISM in our
Galaxy.

CaII absorption has been detected in some Damped Ly$\alpha$ (DLA)
systems (Khare et al. 2004). These systems are believed to be
observed when the line of sight to a quasar passes through
galactic disks.
Evidence for the presence of dust in such systems has been derived
from the comparison of continuum and emission lines between
quasars with DLA and those without; the dust-to-gas ratio has been
found to be 0.05$-$0.20 that of the Galactic for the systems at
relative high redshifts $1.8\le Z_a\le 3.4$ and with column
density $ 1\times 10^{20}\le N_H \le 5\times 10^{21}$~cm$^{-2}$
(Pei, Fall, \& Bechtold 1991). Heavy reddening and CaII absorption
were also reported in two gravitational lensed quasars by lensing
galaxies at intermediate redshifts ($E(B-V)\simeq 0.5$ for APM
08279+5255, Petitjean et al. 2000; $W_{\mathrm CaII\; K}$=5.3\AA~,
E(B$-$V)=1.34 for PMN J0134$-$0931, Hall et al.  2002b). It is
interesting to note that $W_{\mathrm CaII\; K}$/E(B$-$V) measured
in PMN J0134$-$0931 is similar to that in SDSS J2339$-$0912, while
a meaningful comparison is not permitted for APM 08279+5255 since
Petijean et al. (2000) only gave the CaII EWs of two narrow
components.

Strong CaII absorption due to ISM at the quasar redshift is rarely
reported\footnote{Blue-shifted CaII absorption lines have been
detected in Mkn 231 and several other low ionization broad
absorption line quasars, which are considered to be caused in the
dense outflow from the quasars (e.g., Rupke, Veilleux, \& Sanders
2002).}.
Through one side of the Galactic disk, the reduced EW (
$W_{CaII}\times \sin b$) at infinity is 130 m\AA~ at most (??)
according to Smoker et al. (2003). To give arise to the observed
$W_{CaII\; K}=4.2\AA$ in SDSS J2339$-$0912, a nearly edge-on disk
is required. Large N(NaI)/N(CaII) ($>$1) is usually considered as
an evidence for disk material (??). Assuming the NaI absorption
line profile is similar to that of CaII, we obtain an upper limit
on EW(NaI $\lambda$3303) 2.8 \AA~ in the source rest frame
and N(NaI)/N(CaII) $<$ 16. Although this limit is not useful to
constrain the origin of the absorbing gas, future observation may
allow us to do this.

The disk can be associated with the quasar host galaxy or a close
companion of the host. However, the large width of the absorption
line is difficult explain unless the disk gas is severely
disturbed. The warm gas is expected to be highly dispersive. If
the galactic disk is not severely disturbed, the warm gas would
relax onto the disk plane, and its orbit wpuld be almost circular. In
the host galaxy case, the gas velocity would be essentially
perpendicular to the line of sight toward the galactic center, so
that the projected velocity on the line of sight would be very
small. In the companion galaxy case where the line of sight toward
the quasar intersects only one side of the disk, the rotation
speed should be at least 720 km~s$^{-1}$ to produce a CaII profile
of $\sim$ 360 km~s$^{-1}$ FWHM. Following the Tully-Fisher
relation (Pierce \& Tully 1992), the maximum absolute magnitude of
the companion galaxy would be $-24$ mag at B band, more luminous
than the reddened quasar ($B=-23.6$ mag).

We suggest that the absorber is the warm diffuse gas in the
galactic disk of the host galaxy or a close companion of the host,
and that it is highly disturbed by a galaxy collision or even
merger. Disturbed gas has been proposed in the study of the
low-redshift quasar absorption systems in a few quasar-galaxy
pairs, and was considered to be evidence of
a recent merger or an ongoing encounter with a companion(Carilli
\& van Gorkom 1992). 
The detection of blueshifted Na I D absorption lines in the ultraluminous
infrared galaxies (Rupke et al. 2002), which are thought to be galaxies in collision, 
further supports this idea. The large line width and high
column density of CaII presented in SDSS J2339$-$0912 require much
stronger perturbation or even the merging of two galaxies.
 Future high
S/N and resolution spectra in the near-UV/optical can allow us to
precisely resolve the line structure of CaII and other metals,
yielding the chemical abundance and the kinematic structure of the
absorbing gas.

\acknowledgements This work is supported by Chinese NSF grant
NSF10233030, the Bairen Project of CAS, and a key program of the
Chinese Science and Technology Ministry. We thank the anonymous
referee for helpful comments that lead to the improvement of this
paper. This paper has made use of the data from the SDSS and 2MASS
projects.
Funding for the creation and the distribution of the
SDSS Archive has been provided by the Alfred P. Sloan Foundation,
the Participating Institutions, the National Aeronautics and Space
Administration, the National Science Foundation, the U.S.
Department of Energy, the Japanese Monbukagakusho, and the Max
Planck Society. The SDSS is managed by the Astrophysical Research
Consortium (ARC) for the Participating Institutions. The
Participating Institutions are The University of Chicago,
Fermilab, the Institute for Advanced Study, the Japan
Participation Group, The Johns Hopkins University, Los Alamos
National Laboratory, the Max-Planck-Institute for Astronomy
(MPIA), the Max-Planck-Institute for Astrophysics (MPA), New
Mexico State University, Princeton University, the United States
Naval Observatory, and the University of Washington.

\begin{figure}
\epsscale{0.6} \plotone{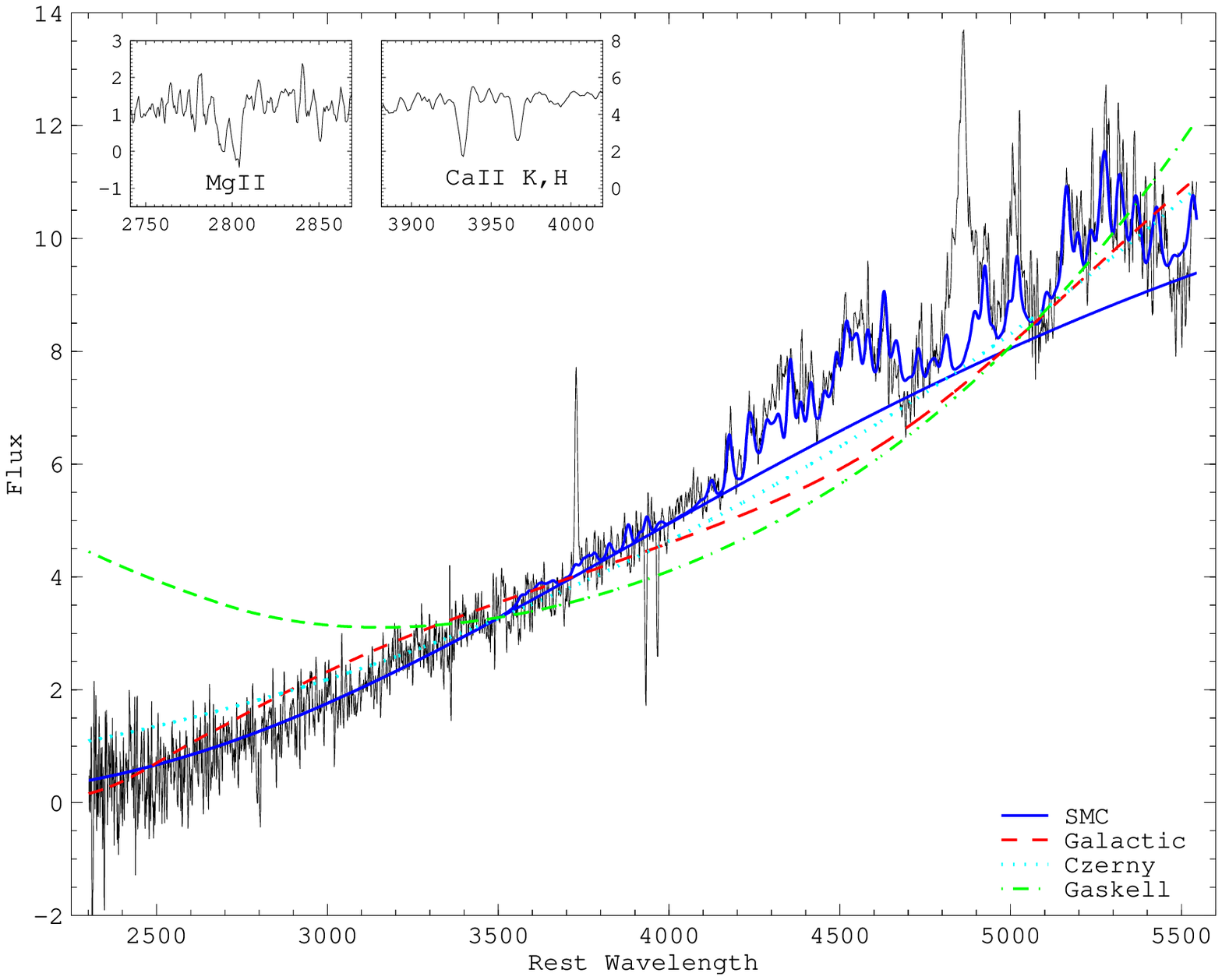} \caption{The observed
spectrum (black thin line) of SDSS J2339-0912 and the modelled
spectrum (blue thick line) of FeII and power-law continuum
reddened by E(B-V)=0.97 for SMC-like extinction. The best fitted
power-laws using extinction curves of SMC, the Galaxy, Czerny et
al. (2004) and Gaskell et al. (2004) are also shown for
comparison. Inserted are enlarged portions of CaII H,K and MgII
absorption lines. The SDSS spectrum has been smoothed with 9
pixels. \label{fig1}}
\end{figure}

\begin{figure}
\epsscale{0.6} \plotone{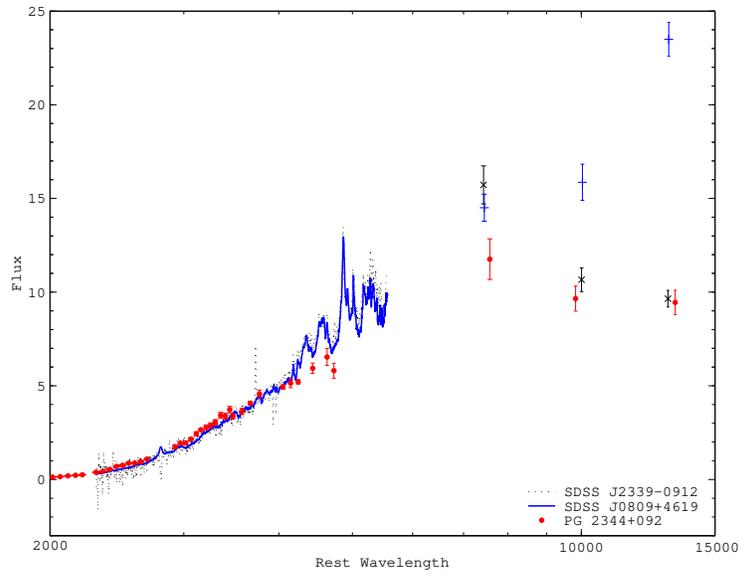} \caption{Comparison of the
SDSS spectrum and near-infrared fluxes of SDSS J2339-0912 (black)
with those of the blue quasar SDSS J0809+4619 (red) reddened by
SMC-like dust of E(B-V)=1.02. Crosses and Circles are photometric
data from the 2MASS. PG 2344+092 (red solid circle; data from
Neugebauer et al. 1987) is also plot.}
\end{figure}


\begin{thebibliography}{}

\bibitem[Abazajian et al.(2004)]{2004AJ....128..502A} Abazajian, K., et
al.\ 2004, \aj, 128, 502

\bibitem[Abazajian et al.(2005)]{} Abazajian, K., et
al.\ 2005, \aj, accepted

\bibitem[Bohlin, Savage, \& Drake(1978)]{1978ApJ...224..132B} Bohlin,
R.~C., Savage, B.~D., \& Drake, J.~F.\ 1978, \apj, 224, 132

\bibitem[Boroson \& Green(1992)]{1992ApJS...80..109B} Boroson,
T.~A., \& Green, R.~F.\ 1992, \apjs, 80, 109 (BG92)

\bibitem[Carilli \& van Gorkom(1992)]{1992ApJ...399..373C} Carilli,
C.~L.~\& van Gorkom, J.~H.\ 1992, \apj, 399, 373

\bibitem[Carrera, Page, \& Mittaz(2004)]{2004A&A...420..163C} Carrera,
F.~J., Page, M.~J., \& Mittaz, J.~P.~D.\ 2004, \aap, 420, 163


\bibitem[Cutri, Nelson, Huchra, \& Smith(2000)]{2000AAS...197.7408C} Cutri,
R.~M., Nelson, B.~O., Huchra, J.~P., \& Smith, P.~S.\ 2000,
Bulletin of the American Astronomical Society, 32, 1520

\bibitem[Czerny, Li, Loska, \& Szczerba(2004)]{2004MNRAS.348L..54C} Czerny,
B., Li, J., Loska, Z., \& Szczerba, R.\ 2004, \mnras, 348, L54

\bibitem[Dong et al.(2005)]{} Dong, X., Zhou, H., Wang, T., Wang, J. Li, C., \& Zhou, Y. \ 2005, \apj,
in press

\bibitem{} Francis, P.J. 1996, Publ. Astron. Soc. Australia, 13, 212

\bibitem[Gaskell et al.(2004)]{2004ApJ...616..147G} Gaskell, C.~M.,
Goosmann, R.~W., Antonucci, R.~R.~J., \& Whysong, D.~H.\ 2004, \apj, 616,
147

\bibitem[Goodrich(1989)]{1989ApJ...342..224G} Goodrich, R.~W.\ 1989, \apj,
342, 224


\bibitem[Hall et al.(2002a)]{2002ApJS..141..267H} Hall, P.~B., et al.\ 2002a,
\apjs, 141, 267

\bibitem[Hall et al.(2002a)]{2002ApJ...575L..51H} Hall, P.~B., Richards,
G.~T., York, D.~G., Keeton, C.~R., Bowen, D.~V., Schneider, D.~P.,
Schlegel, D.~J., \& Brinkmann, J.\ 2002b, \apjl, 575, L51

\bibitem[Hopkins et al.(2004)]{2004AJ....128.1112H} Hopkins, P.~F., et al.\
2004, \aj, 128, 1112

\bibitem[Khare et al.(2004)]{2004astro.ph..8139K} Khare, P., Kulkarni,
V.~P., Lauroesch, J.~T., York, D.~G., Crotts, A.~P.~S., \& Nakamura, O.\
2004, ArXiv Astrophysics e-prints, astro-ph/0408139

\bibitem[Loaring, Page, \& Ramsay(2003)]{2003MNRAS.345..865L} Loaring,
N.~S., Page, M.~J., \& Ramsay, G.\ 2003, \mnras, 345, 865

\bibitem[Maiolino et al.(2001)]{2001A&A...365...28M} Maiolino, R., Marconi,
A., Salvati, M., Risaliti, G., Severgnini, P., Oliva, E., La
Franca, F., \& Vanzi, L.\ 2001, \aap, 365, 28

\bibitem[Neugebauer et al.(1987)]{1987ApJS...63..615N} Neugebauer, G.,
Green, R.~F., Matthews, K., Schmidt, M., Soifer, B.~T., \& Bennett, J.\
1987, \apjs, 63, 615

\bibitem[Pei, Fall, \& Bechtold(1991)]{1991ApJ...378....6P} Pei, Y.~C.,
Fall, S.~M., \& Bechtold, J.\ 1991, \apj, 378, 6

\bibitem[Pierce \& Tully(1992)]{1992ApJ...387...47P} Pierce, M.~J.~\&
Tully, R.~B.\ 1992, \apj, 387, 47

\bibitem[Petitjean, Aracil, Srianand, \& Ibata(2000)]{2000A&A...359..457P}
Petitjean, P., Aracil, B., Srianand, R., \& Ibata, R.\ 2000, \aap,
359, 457

\bibitem[Richards et al.(2003)]{2003AJ....126.1131R} Richards, G.~T., et
al.\ 2003, \aj, 126, 1131

\bibitem[Rupke et al.(2002)]{2002ApJ...570..588R} Rupke, D.~S., Veilleux,
S., \& Sanders, D.~B.\ 2002, \apj, 570, 588

\bibitem[Smoker et al.(2003)]{2003MNRAS.346..119S} Smoker, J.~V., et al.\
2003, \mnras, 346, 119

\bibitem[Wakker \& Mathis(2000)]{2000ApJ...544L.107W} Wakker, B.~P.~\&
Mathis, J.~S.\ 2000, \apjl, 544, L107


\bibitem[Webster et al.(1995)]{1995Natur.375..469W} Webster, R.~L.,
Francis, P.~J., Peterson, B.~A., Drinkwater, M.~J., \& Masci,
F.~J.\ 1995, \nat, 375, 469

\bibitem[Weingartner \& Murray(2002)]{2002ApJ...580...88W} Weingartner,
J.~C., \& Murray, N.\ 2002, \apj, 580, 88

\bibitem[York et al.(2000)]{2000AJ....120.1579Y} York, D.~G., et al.\ 2000,
\aj, 120, 1579

\bibitem[Zuo et al.(1997)]{1997ApJ...477..568Z} Zuo, L., Beaver, E.~A.,
Burbidge, E.~M., Cohen, R.~D., Junkkarinen, V.~T., \& Lyons,
R.~W.\ 1997, \apj, 477, 568
\end{thebibliography}
\end{document}